

\input harvmac
\input epsf

\Title{\vbox{\baselineskip12pt\hbox{EFI-93-73}\hbox{hep-th/9312127}}}
{\vbox{\centerline{Black Hole Evaporation
along Macroscopic Strings}}}

\centerline{\it Albion Lawrence\footnote{$\dagger$}{Supported
in part by an NSF Graduate Fellowship.}
and Emil Martinec\footnote{*}{Supported
in part by Dept. of Energy grant DEFG02-90ER-40560.}}
\smallskip
\centerline{Enrico Fermi Inst. and Dept. of Physics}
\centerline{University of Chicago, Chicago, IL  60637}
\vskip .2 in

\noindent
We develop the quantization of a macroscopic string which extends radially
from a Schwarzschild black hole.  The Hawking process excites
a thermal bath of string modes that causes the black hole to lose mass.
The resulting typical string configuration is a random walk in the angular
coordinates.
We show that the energy flux
in string excitations is approximately that of
spacetime field modes.
\vskip 1.5cm
\centerline{{\it submitted to Phys. Rev. D}}

\Date{1/94}


\def\anp{Ann. Phys.}
\def\np{Nucl. Phys.}
\def\pl{Phys. Lett.}
\def\pr{Phys. Rev.}
\def\prl{Phys. Rev. Lett.}

\def\journal#1&#2(#3) {\unskip, \sl #1\ \bf #2 \rm(19#3) }
\def\andjournal#1&#2(#3){\sl #1~\bf #2 \rm (19#3) }


\def\bra#1{{\langle #1 |}}
\def\ket#1{{| #1 \rangle}}
\def\vev#1{{\langle #1 \rangle}}
\def\sig{$\sigma$} 
\def\vol2{d^{2}\sigma} 
\def\rst{r_{\ast}}  
\def\frac#1#2{{#1 \over #2}}
\def\hab{h^{\alpha \beta}} 

\def\sst{\scriptscriptstyle}
\def\d{\partial}
\def\dep{\partial_{+}}
\def\dem{\partial_{-}}
\def\sp{\sigma^{+}} 
\def\sm{\sigma^{-}}
\def\scrip{{\cal I}^{+}} 
\def\scrim{{\cal I}^{-}} 
\def\qbrs{Q_{{\sst \rm BRST}}} 

\def\rh{$r_{h}$} 
\def\apr{$\alpha'$} 
\def\MM{{\cal M}}
\def\KK{{\cal K}}
\def\PP{{\cal P}}

\def\ie{{\it i.e.}}

\def\et{{\it et.al.}}
\def\etc{{\it etc.}}
\def\eg{{\it e.g.}}

\newsec{Introduction}

The collision of quantum mechanics and black hole physics has
captured the interest of the theoretical physics community.
The discovery of two-dimensional model systems \ref\cghs{C. Callan,
S. Giddings, J. Harvey and A. Strominger\journal \pr&D45 (92)
1005.} sharing many
of the features of the Hawking problem has spurred the hope that one
might learn something about information storage and retrieval
in black holes, and perhaps about some essential features of quantum gravity.
Before this hope evaporates (with or without a remnant)
we would like to point out another context in which such two-dimensional
systems crop up: the quantization of the collective modes of a macroscopic
string trapped by a black hole.

Often one ties two-dimensional model systems
to four-dimensional physics when, at the core of a spherically symmetric
object, all but S-wave modes of the four-dimensional system are
effectively massive~\ref\callanrubakov{C. Callan\journal\np&B212 (83) 391;
V. Rubakov\journal {\sl JETP}&33 (81) 644.} \ref\strom{M. Alford
and A. Strominger\journal \prl&69 (92) 563.}
and may be integrated out.
For a macroscopic string, a somewhat different mechanism occurs.
The internal excitations of the string core are irrelevant
to the physics of the Nambu-Goldstone transverse oscillation modes
(for a macroscopic fundamental string, there are no such core excitations).
The dominant dynamical effects come from
the projection of the spacetime geometry
onto the world sheet of the propagating string.
When the string is trapped by the event horizon of a black hole,
the event horizon is projected onto the world sheet.  Hence one expects
Hawking radiation of transverse string oscillations to occur
upon quantization.
Since the essence of the Hawking
paradox is whether information follows energy, it would seem that one
ought to understand it in this somewhat simpler
two-dimensional context.
Any remnant scenario would entail
a huge expansion of the string density of states, since the remnant
would be threaded by the string (although such remnants
might be hard to produce, as may be the case for recently
considered species of remnant~\ref\banks{D. Garfinkle,
S. Giddings and A. Strominger, {\it Entropy in black hole
pair production}, Santa Barbara preprint
UCSBTH-93-17, gr-qc/9306023.}).  If black holes violate
quantum mechanical unitarity, we will have to face
the problem already in the quantization of a single string.

Hawking radiation along the string leads to a number of interesting
questions, of which we shall address two,
which involve effects of the radiation on the string
and its back-reaction on spacetime.  We will show that the thermal bath
of emitted string modes causes the mean square
transverse extent of the string to grow linearly
with its length (or some other infrared cutoff) $L$
$$
  \vev{:(\xi_\perp)^2:}\sim \frac{\ell_s^2 L}{r_h}\ ,
$$
and that the power radiated in the string Hawking process is
\eqn\stringrad{
  \left(\frac{\delta M_{\rm bh}}{\delta t}\right)_{\rm string~modes}
	\sim \frac{\hbar c^2}{r_h^{2}}\ ,
}
where $r_{h}$ is the horizon radius.
This is to be compared with, \eg, the total radiated power
a spacetime massless scalar field in the presence of a black hole,
which is the same within numerical factors.

We begin in section 2 with the analysis of the
small fluctuations of a single straight, infinitely long,
bosonic string in a
Schwarzschild black hole background in the critical
dimension of string theory by expanding the $\sigma$-model
action to quadratic order in Riemannian normal coordinates.  In section 3
we discuss the linearized classical equations of motion and show that
the physical
solutions are stable to this order.
In section 4 we quantize the transverse fluctuations in a physical
gauge, demonstrating both through this gauge fixing and through
path integral techniques in conformal gauge
that there are~($d_{cr}$--2) independent
physical coordinates in the critical dimension.  We choose the
Unruh vacuum as the string state and construct the Bogolubov
transformations used to describe Hawking radiation.
Those who are familiar with this material may wish to skip directly
to section 5, where we compute some
observable manifestations of the Hawking process: the
mean square deviation from the equilibrium position and the radiation of
spacetime energy onto the string, both characteristic of a string
thermally excited at the Hawking temperature.

\newsec{The string action in a Schwarzschild background}

\subsec{The background and the \sig-model action}

	The metric of the Schwarzschild black hole in D
dimensions is derivable in the same fashion as the 4-dimensional
case, and differs only in the power of the radial coordinate dependence
of the metric for the $r-t$ plane.
(see for example~\ref\mype{R. Myers and M. Perry
\journal\anp&172 (86) 304}):
\eqn\metric{
	ds^2 = - \left( 1 - \frac{C}{r^{D-3}} \right) dt^2
	+ \left( 1 - \frac{C}{r^{D-3}} \right)^{-1} dr^{2} +
	r^{2} d\Omega^{2}\ .
}
Here $d\Omega^{2}$ is the metric for the D-2 sphere;
$C$ is a constant of integration arising from solving Einstein's equations
and is related to the mass via \mype
\eqn\mass{
	M = \frac{C (D-2) A_{D-2}}{16 \pi G}\ ,
}
where $A_{D-2}$ is the area of a unit D-2 sphere and G is the D-dimensional
generalization of Newton's constant, defined via the Einstein action
with dimensions $[{\rm length}]^{D-2}$ ($\hbar~=~c~=~1$).
The radius of the horizon is of course $r_{h}=C^{1/(D-3)}$.

\goodbreak\midinsert\centerline{\epsfxsize=6truein\epsfbox{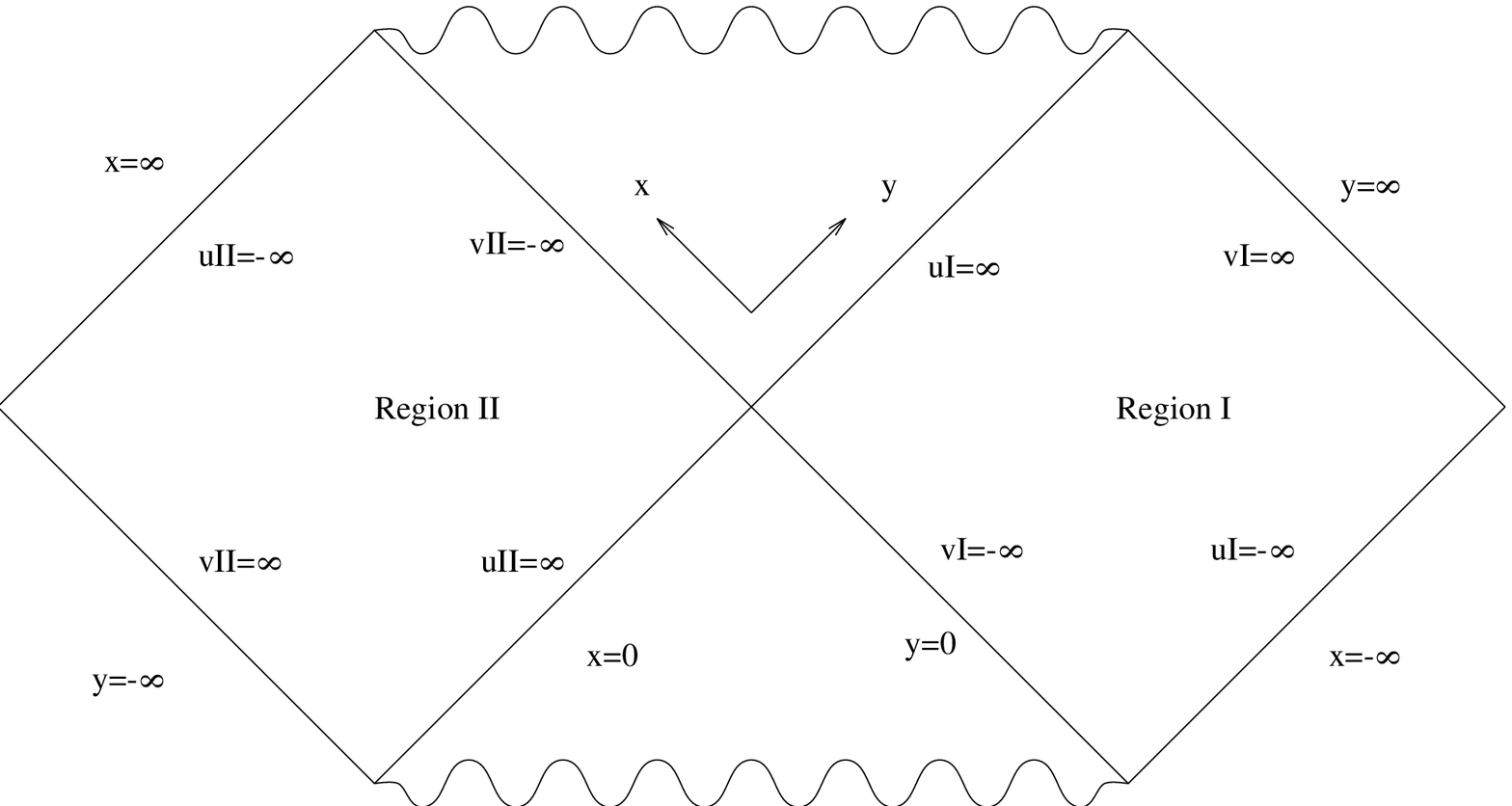}}
\narrower\narrower\noindent{\footnotefont {\bf Figure 1}
Penrose diagram for the x-y plane of the extended Kruskal manifold.\par}
\endinsert

	In this paper we will be switching between Schwarzschild null
coordinates and Kruskal null coordinates.  The former are found by
solving for $ds^{2} = 0$ in the r-t plane, and are
\eqn\schnul{
	\eqalign{
	u & = t - \rst \cr
	v & = t + \rst\ ,}
}
where $\rst$ is the D-dimensional
generalization of the Regge-Wheeler tortoise coordinate
\ref\wald{R. M. Wald, {\it{General Relativity}}, (U. of
Chicago Press, Chicago, 1984)}
\eqn\tort{
	\rst = \int^{r} \frac{dr}{1 - \left( \frac{r_{h}}{r} \right)^{D-3}}
	= r + r_{h}\int^{r/r_{h}} \frac{dz}{z^{D-3} - 1}\ .
}
The indefinite integral may be computed exactly by
expanding the denominator in partial fractions, with the
result:
\eqn\inttort{
	\rst = r + \frac{r_{h}}{(D-3)} \sum_{k=0}^{D-4}
	\ln\left(\frac{r}{r_{h}} - e^{i2\pi k/(D-3)}\right)\ .
}
The metric may then be written as:
\eqn\scnmet{
	ds^{2} = -\left[ 1 - \left(\frac{r_{h}}{r}\right)^{D-3}
		\right] du dv + r^{2}d\Omega^{2}\ .
}
Kruskal null coordinates are used here in dimensionless form,
\eqn\krusk{
	\eqalign{
	x & = - e^{-(\frac{D-3}{2r_{h}})u} \cr
	y & = \hbox{ } e^{(\frac{D-3}{2r_{h}})v}\ ;
}}
the $r-t$ metric is, in these coordinates,
\eqn\krmet{
	\eqalign{
	ds^{2} & = - 2 f(x,y)\ dx dy \cr
		& = - \frac{4 r_{h}^{2}}{(D-3)^{2} xy}\
		\left[1-\left(\frac{r_{h}}{r}\right)^{D-3}\right]\ dx dy\ .}
}
Figure 1 shows the standard Penrose diagram for the ``r-t'' or
``x-y'' plane of the black hole, labelled with our notational
conventions.

	We wish to study an infinitely long string fluctuating in
this background; we will assume that the string and its fluctuations
have a negligible effect on the spacetime.  We will also
assume that the black hole has a large mass
or equivalently a large horizon radius \rh,  which will
allow us to expand the action out in inverse powers
of \rh; in addition, since the Hawking temperature goes as the inverse of
the horizon radius, we may treat adiabatically the change in the background
metric due to the shrinking of the black hole.
The action may be written as the \sig-model
action\foot{We adopt the following conventions.
For vector and tensor indices: latin
indices in the first half of the alphabet ($a,b$...)
are worldsheet tangent space indices,  greek indices in the first half of
the alphabet ($\alpha$,~$\beta$,...) are worldsheet coordinate
indices,  latin indices in the last half of the alphabet ($m$,$n$,...) are
spacetime tangent space indices, and greek indices in the last half
of the alphabet ($\mu,~\nu$...) are spacetime coordinate indices.
The flat space
metric signature is $\eta_{\alpha\beta}=(-,+)$ on the worldsheet and
$\eta_{\mu\nu}=(-,+,...,+)$ in space time.}
\eqn\sigact{
	S = \frac{1}{4\pi\alpha'} \int d^{2}\sigma\sqrt{h} \left\{
		h^{\alpha\beta}
		g_{\mu\nu} \d_{\alpha} X^{\mu} \d_{\beta} X^{\nu}
		\right\} + S_{{\rm ghosts}}\ .
}
The ghost action will arise from fixing to conformal gauge with some fiducial
metric ${\hat h}_{\alpha\beta}$ multiplied by an arbitrary conformal factor
$\rho$. The ghost action will be the same as that of a string in Minkowski
spacetime; it arises purely from the worldsheet geometry, and will
depend explicitly on the fiducial metic and implicitly (through
ultraviolet regulation) on $\rho$.
Note that since the space-time metric is Ricci-flat, this
theory is free of the
conformal anomaly in 26 dimensions
\ref\callan{C.G. Callan, D. Friedan,
E.J. Martinec and M.J. Perry \journal\np&B262 (85) 593}.
The generalization to
the supersymmetric \sig-model in this background is straightforward.

\subsec{Normal coordinate expansion}

	We would like to examine small fluctuations
around an equilibrium position of the string, \ie\ about some solution
to the classical equations of motion; thus, we want
some sort of Taylor expansion of $X^{\mu}$ around
the equlibrium configuration $X^{\mu}_{\rm cl}$.
We find that, in Kruskal coordinates,
\eqn\trivsoln{
	\eqalign{
	X^{x}_{\rm cl} & = \sigma^{-} \cr
	X^{y}_{\rm cl} & = \sigma^{+} \cr
	X^{\theta_{k}}_{\rm cl} & = \frac{\pi}{2}
}}
is the coordinate embedding
for an infinitely long string in conformal gauge.
This embedding defines
a one-to-one mapping between the $x-y$ plane of spacetime and the
classical worldsheet of the string.  We may also pull back the metric
for the spacetime $x-y$ plane coordinates onto worldsheet, fixing
the fiducial metric as:
\eqn\setfiducial{
	\hat{h}_{\alpha \beta} d\sigma^{\alpha} d\sigma^{\beta}
	= -2 f d\sigma^{-} d\sigma^{+}\ .
}

Riemannian normal coordinates are a standard and natural
way to carry out a Taylor expansion of the \sig-model action
in a manifestly coordinate invariant
fashion (see~\ref\alvgaum{L. Alvarez-Gaum\'e, D.Z Freedman
and S. Mukhi \journal\anp&143 (81) 85}
for an explanation of this expansion in the
bosonic and supersymmetric \sig-models).
In particular, it is an expansion in derivatives of the
metric;  one finds that a term in the action of order $n$
in small fluctuations is accompanied by geometric coefficients
(derivatives of and combinations of derivatives of the Riemann tensor)
containing $n$ derivatives of the metric.
Since the only dimensionful quantity in
the metric is $r_{h}$, we may regard this expansion as an
expansion in the fluctuation size divided by \rh; then the
expansion is good if the fluctuations are small on the
scale of the black hole or if one is in the asymptotic region.
Classically, this means simply that we
wish to perturb the string (shake it) with small amplitude.
Quantum mechanically, the strength of worldsheet fluctuations is
given by $1/\ell_{s}^{2}$ which appears in front of the nonlinear
\sig-model action; thus, perturbations to expectation values coming from
nonlinear terms will be of higher order in $\ell_{s}/r_{h}$.

Following~\alvgaum, let us expand the coordinate fields and the spacetime
metric in normal
coordinates, which are contravariant
vectors $\xi^{\mu}$ in spacetime.
We may rewrite these as
spacetime tangent frame quantities,
\eqn\tncoords{
	\xi^{m} = e^{m}_{\mu}(X_{\rm cl}) \xi^{\mu}\ .
}
where $e^{m}_{\mu}(X_{\rm cl})$ is the {vielbein} field of the spacetime
manifold at the spacetime coordinate $X_{\rm cl}$.  By working with these
{vielbein } fields we will have a more standard kinetic term and
thus a more standard propagator.
The final form of the action for small coordinate
fluctuations to quadratic order is:
\eqn\quadact{
	\eqalign{
	S = - \frac{1}{2} \int \vol2 \sqrt{h} \biggl\{
	& \eta_{mn} \hab (D_{\alpha} \xi)^{m} (D_{\beta} \xi)^{n} \cr
	& R_{\mu m n \nu} \xi^{m} \xi^{n} \hab \d_{\alpha}X_{\rm cl}^{\mu}
		\d_{\beta}X_{\rm cl}^{\nu} \biggr\} + S_{{\rm ghosts}}\ .
}}
Here $\eta$ is a flat Minkowski metric for the tangent frames, and
the derivatives $D$ are defined by
\eqn\tancovder{
	(D_{\alpha}\xi)^{m} = \d_{\alpha} \xi^{m} +
		\omega_{\mu}^{mn} \d_{\alpha} X_{\rm cl}^{\mu} \xi_{n}\ ,
}
where $\omega_{\mu}^{mn}$ is the spacetime spin connection.

	The {vielbein} components are:
\eqn\vb{
	\eqalign{
	e_{x}^{0} & = \sqrt{\frac{f}{2}} = e_{y}^{0} = e_{y}^{1} \cr
	e_{x}^{1} & = - \sqrt{\frac{f}{2}} \cr
	e_{\theta_{k}}^{m} & = r(x,y) \sin\theta_{D-2} \cdot \cdot \cdot
		\sin \theta_{k+1}
	\qquad \hbox{for 2 $\leq$ k $\leq$ $D-1$}
}}
($r$ is defined implicitly via \schnul\ - \krusk).
We may use these to calculate $\omega_{\mu}^{mn}$ and $R_{\mu\nu\rho\sigma}$.
The explicit form of \quadact\ in Kruskal coordinates is found to be:
\eqn\quadcoord{
	\eqalign{
	S = - \half \int \vol2 \Biggl\{ &
	2 \eta_{ab} \dem \xi^{a} \dep \xi^{b} \cr
	+ & \frac{1}{f} \left[ \xi^{1} (\dem f \dep - \dep f \dem) \xi^{0}
		- \xi^{0} (\dem f \dep - \dep f \dem) \xi^{1} \right] \cr
	+ & \left( \frac{\dem f \dep f - \half f \dem\dep f}{f^{2}} \right)
	\left( - \xi^{0}\xi^{0} + \xi^{1}\xi^{1} \right) \cr
	+ & 2 \frac{\dem\dep r}{r} \sum_{k=2}^{D-1}
	\xi^{k} \xi^{k}\Biggr\} \cr
	+ & S_{{\rm ghosts}}\ .
}}
(In fact, if we let \hbox{$\sigma^{-}. \sigma^{+} = X_{{\sst cl}}^u,
X_{{\sst cl}}^v$},
\hbox{$ f = -\half (1 - (r_h/r)^{D-3})$}, then \quadcoord\ is the
action in Schwartzchild null coordinates as well.)
Note that this action has the standard kinetic term for the transverse
coordinates; in general, the curvature terms will be some finite
r-dependent piece (vanishing as a power of $r_h/r$ asymptotically)
times a factor of $1/r_{h}^{2}$.  Hence the kinetic term dominates
the action far from the horizon.  On the other hand, near the horizon
the modes relevant to Hawking radiation are highly blue-shifted
and also dominate the curvature terms.

The world sheet stress tensor comes from varying \sigact\ with respect to
$h_{\alpha\beta}$:
\eqn\fullstress{
	\eqalign{
	T_{\pm\pm} & = g_{\mu\nu}\d_{\pm}X^{\mu}\d_{\pm}X^{\nu} \cr
	T_{+-} & = 0 \qquad \hbox{(classically)}\ .
}}
Using \alvgaum\ to expand this to quadratic order, we find that:
\eqn\stress{
	\eqalign{
	T_{--} & = 2 g_{xy} \dem X^{x}_{{\rm cl}} D_{-}\xi^{y} +
	2 g_{xy} D_{-}\xi^{x}D_{-}\xi^{y}
	+ D_{-}\xi^{k}\cdot D_{-}\xi^{k} + R_{x\kappa\lambda x}\xi^{\kappa}
		\xi^{\lambda} \cr
	T_{++} & = 2 g_{xy}\dep X^{y}_{{\rm cl}} D_{+}\xi^{x} +
	2 g_{xy} D_{+}\xi^{x}D_{+}\xi^{y}
	+ D_{+}\xi^{k}\cdot D_{+}\xi^{k} + R_{y\kappa\lambda y}\xi^{\kappa}
		\xi^{\lambda}\ ,
}}
where $k$ is summed over the angular variables.  Here we
have written the longitudinal modes as coordinate vector fields rather than
as tangent frame vectors, as we will find this form to be
easier to use in explicitly solving the constraints.  For
the Schwarzschild metric, the curvature
term couples to the longitudinal modes only, and the explicit
form of $T_{\pm\pm}$ is
\eqn\stressexp{
	\eqalign{
	T_{--} & = -2f \d_{-}\xi^{y}
	-2f\left(\d_{-}\xi^{x}+\frac{\d_{-}f}{f}\xi^{x}\right)\d_{-}\xi^{y}
	\cr
	& + \frac{2 (D-2) r_{h}^{2}}
	{(D-3)^{3}(\sigma^{+}\sigma^{-})^{2}} \frac{D-2}{D-3}
	\left(\frac{r_{h}}{r}\right)^{D-1}
	\left[1 - \left(\frac{r_{h}}{r}\right)^{D-3}\right]^{2}
	\left(\xi^{y}\right)^2 \cr
	& + \d_{-}\xi^{k}\cdot\d_{-}\xi^{k} \cr
	T_{++} & = T_{--} (\d_{-} \leftrightarrow \d_{-},
	x \leftrightarrow y)\ .
}}

\newsec{Classical physics}

\subsec{Mode solutions}

The equations of motion for the angular coordinates are
\eqn\angcleqn{
	\dep\dem\xi^{k} - \frac{\dep\dem r}{r} \xi^{k} = 0,
	\qquad 2 \geq k \geq D-1\ .
}
One may write the second (curvature) term more explicitly:
\eqn\curvepert{
	\frac{\d_{+}\d_{-}r}{r} =
	\left(\frac{r_{h}}{r}\right)^{D-1}
	\frac{1}{(D-3)\sigma^{+}\sigma^{-}}
	\left[1 - \left(\frac{r_{h}}{r}\right)^{D-3}
	\right]\ ,
}
again with $r=r(x=\sigma^{-},y=\sigma^{+})$.  It is easy to see that
\curvepert\ vanishes for large $r$.  For $r$ close to $r_h$, we may
substitute in \krusk\ and \schnul, and use \inttort,
to find that the zero in $\sigma^{+}\sigma^{-}$ cancels the zero in
$[1 - (r_{h}/r)^{D-3}]$, and that the result is a finite number.
As we shall see, all the modes will be blueshifted infinitely at the
horizon, so that the kinetic term will then dominate close to the horizon.
Thus we will, after ensuring stability, use the
zeroth-order approximation that the transverse modes satisfy the 2-d
massless Klein-Gordon equation for the purpose of computing
Bogolubov transformations between mode bases.
The inaccuracy of this procedure lies in the fact that there is
mixing between left- and right-moving modes due to scattering
off the background curvature.  Hence the Bogolubov coefficients
will mix left- and right-moving creation and annihilation operators.

	We were	unable to solve~\angcleqn\ exactly,
but one can find solutions that
work asymptotically as \hbox{$r_{*}\rightarrow-\infty$}
(\hbox{$r\rightarrow r_{h}$}) and as \hbox{$r,\rst\rightarrow\infty$}.
To do this, transform the variables as follows:
\eqn\excoords{
	\eqalign{
	q & = \sqrt{- \sigma^{-} \sigma^{+}} \cr
	s & = \sqrt{- \frac{\sigma^{+}}{\sigma^{-}}}\ .
}}
{}From \krusk~one can see that
\eqn\qsasrt{
	\eqalign{
	q & = e^{\frac{D-3}{2r_{h}} \rst(x=\sigma^{-},y=\sigma^{+})} \cr
	s & = e^{\frac{D-3}{2r_{h}} t(x=\sigma^{-}, y=\sigma^{+})}\ .
}}
Note that these coordinates are only defined in one or the other regions of
Figure I which are outside the horizon; we will define an appropriate
analytic continuation below.

	In these coordinates the equations of motion separate:
\eqn\sepang{
	s^{2} \biggl(\frac{\d^{2}}{\d s^{2}} - \frac{1}{s} \frac{\d}{\d s}
		\biggr) \xi^{k}
	= q^{2} \biggl(\frac{\d^{2}}{\d q^{2}} - \frac{1}{q} \frac{\d}{\d q}
	\biggr) \xi^{k}
	-4\frac{1}{D-3}\biggl(\frac{r_{h}}{r}\biggr)^{D-1}
	\Biggl[ 1 - \biggl( \frac{r_{h}}{r} \biggr)^{D-3} \Biggr] \xi^{k}\ .
}
One finds that at null infinity and in towards the horizon, the last term
vanishes.  Thus, in this asymptotic sense, the wave
equation for the transverse coordinates is
\eqn\trapproxsoln{
	s^{2} \biggl(\frac{\d^{2}}{\d s^{2}} - \frac{1}{s} \frac{\d}{\d s}
		\biggr) \xi^{k} =
	q^{2} \biggl(\frac{\d^{2}}{\d q^{2}} - \frac{1}{q} \frac{\d}{\d q}
		\biggr) \xi^{k} = - \omega^{2} \xi^{k}\ ,
}
where $\omega$ should be real for stable solutions.  Solutions
to this approximate equation are
\eqn\exsol{
	\xi^{k} = s^{i \omega} q^{\pm i \omega}
}
for any $\omega$.  Translated back into
Kruskal coordinates, the solutions become
\eqn\krsoln{
	\xi^{k} = (\sigma^{-})^{\pm i \omega} \qquad\hbox{ or } \qquad
	(\sigma^{+})^{\pm i \omega}\ .
}

	These mode solutions are standard in Hawking radiation calculations
\ref\unruh{W.G. Unruh \journal\pr&D14 (76) 870}
\ref\birrell{N.D. Birrell and P.C.W. Davies, {\it Quantum Fields in
Curved Space}, (Cambridge University Press, Cambridge, 1982)},
and following
Unruh and Birrell and Davies, we put the
logarithmic branch cut for \krsoln~in the upper half plane by making the
choice
\eqn\branch{
	-1 = e^{- i \pi}\ .
}
With this choice, the solutions in \krsoln~are analytic
in the lower half plane of $x$,
and thus their Fourier transforms vanish for $\omega < 0$, so that the
modes shown are positive frequency in $\sigma^{\pm}$ for all $\omega$.
Note also that
for $r \rightarrow \infty$, these look like plane wave solutions
in Schwarzchild coordinates.

	The equations of motion for the longitudinal modes are most
easily solved in their coordinate vector form.  The equations are:
\eqn\longeom{
	\eqalign{
	\d_{+}\d_{-} \xi^{x} + \frac{\d_{-}f}{f} \d_{+} \xi^{x} & = 0 \cr
	\d_{+}\d_{-} \xi^{y} + \frac{\d_{+}f}{f} \d_{-} \xi^{y} & = 0\ ,
}}
with the solutions
\eqn\longsoln{
	\eqalign{
	\xi^{x} & = \phi^{x}(\sigma^{-}) + \int^{\sigma^{+}}_{-\infty}
			d\sigma^{+}{}'
		\frac{\psi^{x}(\sigma^{+}{}')}
		{f(\sigma^{-},\sigma^{+}{}')} \cr
	\xi^{y} & = \phi^{y}(\sigma^{+}) + \int^{\sigma^{-}}_{-\infty}
			d\sigma^{-}{}'
	   \frac{\psi^{y}(\sigma^{-}{}')}{f(\sigma^{-}{}',\sigma^{+})}\ ,
}}
where $\phi^{x,y}$ and $\psi^{x,y}$ are arbitrary functions.

	The actual classical solutions to this equation should include
the satisfaction of the Virasoro constraints.  If the gradients of the
coordinate fluctuation fields are moderate (as they are asymptotically),
the constraints will
be self-consistently solved by equating the linear term in longitudinal
modes to the quadratic term in transverse modes; the
terms quadratic in longitudinal modes will be of fourth order in
the fluctuation size and should be dealt with in the next order of the
calculation.  Thus we find that in the asymptotic region where
$\xi^{k}~\sim~(\sigma^{+})^{i\omega}$,
$$\eqalign{\psi^{x}\sim &(\sigma^{+})^{2i\omega-2}\cr
  \xi^{x}\sim&\sigma^{-}(\sigma^{+})^{2i\omega}\quad,
	\qquad r\rightarrow\infty\quad.}$$
As $r$ approaches \rh, however,
$f\d\xi^{x}$ will become large and the term in~\stress\ which is second order
in derivatives of longitudinal fluctuations will become even larger and will
become the dominant contribution to the longitudinal stress tensor.
Thus, near the horizon we should equate this quadratic term with
$\d_{\pm}\xi^{k}\d_{\pm}\xi^{k}$.  Specifically,
the piece of the stress tensor quadratic in derivatives of longitudinal
fluctuations is
\eqn\horcon{
	f\dep\xi^{x}(\dep\xi^{y} + \frac{\dep f}{f}\xi^{y})
	= \psi^{x}\left(\dep\phi^{y}
	-\int^{\sigma^{-}} d\sigma^{-'}\frac{\psi^{y}}{f}
		\frac{\dep f}{f} + \frac{\dep f}{f} \xi^{y}\right)\ .
}
Now
\eqn\logdmet{
	\frac{\d_{\pm}f}{f} = - \frac{1}{\sigma^{\pm}}
		\left[ 1 - \left(\frac{r_{h}}{r}\right)^{D-2}\right]\ .
}
As \hbox{$r\rightarrow r_{h}$}, we can find by writing out $\sigma^{\pm}$
in terms of $r$ and $t$ that $\d_{\pm}f/f$ vanishes as
$\sqrt{r~-~r_{h}}$ as we approach the horizon; thus in this limit,
\eqn\horconas{
	f\dep\xi^{x}\left(\dep + \frac{\dep f}{f}\right)\xi^{y}
	\sim \psi^{x}\dep\phi^{y}\ ,
}
and similarly for \hbox{$x\leftrightarrow y$},
\hbox{$\sigma^{+}\leftrightarrow\sigma^{-}$}.  Matching these forms to
$\d_{\pm}\xi^{k}\d_{\pm}\xi^{k}~\sim~(\sigma^{\pm})^{2i\omega-2}$,
we can solve the constraints to lowest order by setting
\eqn\horsat{
	\eqalign{
	\psi^{x,y} & \sim (\sigma^{+,-})^{i\omega - 1} \cr
	\phi^{x,y} & \sim (\sigma^{-,+})^{i\omega}\qquad r\rightarrow r_{h}\ .
}}
Note that with this solution the term quadratic in the derivatives
of longitudinal fluctuations
indeed diverges faster than both the term linear in the derivatives
and the non-derivative quadratic term.

\subsec{Classical linear stability}

	Before we quantize the quadratic theory~\quadcoord, we need to
know whether it is classically consistent -- that is, if small fluctuations
stay small over time.  We should note that simply looking at the size
of $\xi^{\mu}$ over space and time is incorrect since the metric is
invariably singular in some region of spacetime for whatever coordinate
system we might choose -- in $H^{\pm}$ for Schwarzschild coordinates
and in $\cal{I}^{\pm}$ for Kruskal coordinates.
The natural quantity to look at is the scalar product of the fluctuations,
\eqn\norm{
	\eqalign{
	\| \xi \|^2 & = g_{\mu \nu} \xi^{\mu} \xi^{\nu} \cr
			& = 2 g_{xy} \xi^{x} \xi^{y} +
			\delta_{kl} \xi^{k} \xi^{l}
			\qquad \hbox{ (k,l $\geq$ 2)}\ .
}}

	For the transverse fluctuations, in the absence of
exact solutions to \sepang, we shall try to cast the
equations of motion into a Sturm-Liouville form
where stability is obvious.   If we make an additional change of
variables
\eqn\baktobef{
	\eqalign{
	R & = \ln q = \frac{D-3}{2 r_{h}} \rst \cr
	T & = \ln s = \frac{D-3}{2 r_{h}} t\ ,
}}
then the equations of motion become
\eqn\schang{
	\frac{\d^{2}}{\d T^{2}} \xi^{k} - \frac{\d^{2}}{\d R^{2}} \xi^{k}
	+ g(R) \xi^{k} = 0\ ,
}
where
\eqn\massterm{
	g(R) = \frac{4}{D-3} \left(\frac{r_{h}}{r}\right)^{D-1}
	\biggl[1 - \left(\frac{r_{h}}{r}\right)^{D-3}\biggr]\ ,
}
with $r$ used as an implicit function of $R$.  Roughly, \schang~is the
Klein-Gordon equation with a spatially varying mass term.  Since the
``mass squared'' $g(R)$ is always positive, the
normalizable transverse solutions are stable.

	Knowing this, we may use $(\sigma^{\pm})^{i\omega}$ for
real $\omega$ as a complete set of asymptotic solutions for
the transverse equations of motion, and we may then use the
results of the previous subsection to examine the form
of the longitudinal solutions in the asymptotic regions.
For the $r~\rightarrow~\infty$ region,
\eqn\asylong{
	g_{xy} = f \sim \frac{1}{\sigma^{+}\sigma^{-}}\ , \qquad
	\xi^{x} \sim \sigma^{-}(\sigma^{+})^{2i\omega}\ , \qquad
	\xi^{y} \sim \sigma^{+}(\sigma^{-})^{2i\omega}\,
}
so that the scalar product of the longitudinal fluctuations is
$$
	g_{xy}\xi^{x}\xi^{y} \sim (\sigma^{+}\sigma^{-})^{2i\omega}\ .
$$
For the $r \rightarrow r_{h}$ region, f is a regular, finite
quantity, the zero in $1/(\sigma^{+}\sigma^{-})$ cancelling the zero in
$[1 - (r/r_{h})^{D-3}]$.  The form of $\xi^{x,y}$ near the horizon will
be
\eqn\horizonmodes{
	\xi^{x} \sim (\sigma^{+})^{i\omega} +
		\int^{\sigma^{-}}
		d\sigma^{-'} \frac{(\sigma^{-'})^{i\omega-1}}{f} \ ,
}
and similarly for
$x\leftrightarrow y$, $\sigma^{+}\leftrightarrow\sigma^{-}$.
Naive power counting tells us that the second term should be
regular near the horizon, so that $g_{xy}\xi^{x}\xi^{y}$ is
also regular near the horizon.

\newsec{Quantization}

	In this section we would like to address two separate issues:
the imposition of the Virasoro constraints on the theory defined by
\quadcoord, and the problems of selecting physically motivated ``in'' and
``out'' states.  We will first discuss in general the
conformal invariance of the system, and argue that the angular
fluctuations are a good representation of the physical Hilbert space of
the string by showing that the partition function depends only
on the determinants of the transverse fluctuation operators.
We will then discuss the possible definitions of the positive
frequency modes of the transverse fluctuations
and construct the desired vacuum states and their
Bogolubov transformations.  Finally, we shall argue that the Hawking
radiation does not include unphysical modes if the BRST cohomology theorem
holds for this system.

\subsec{Path integral quantization}

	Now the full nonlinear \sig-model action
\sigact\ satisfies the $\beta$-function
equations of \callan\ in the
standard critical dimensions to first order in \apr,
since the metric coupling satisfies Einstein's equations. Thus we
should be able to use the conformal invariance to decouple
the negative-norm states in the Hilbert space of coordinates plus ghosts,
and we expect that in principle one should be able to embark upon a
standard covariant quantization program by imposing the Virasoro
conditions or the corresponding BRST condition.  We have not been able
to successfully use this formalism to prove a no-ghost theorem
due to the subtleties of constructing the vacuum, which involves
correlating the fluctuations of the string
near the horizon and in the asymptotic region.

	If the spectrum is free of
negative norm states and conformal invariance may be imposed
on the quantum level, then the only physical degrees
of freedom are the transverse oscillations of the string, as
in flat space.  We will support this supposition
by showing that the
partition function depends only on the determinants of the
transverse fluctuations;
upon integrating out the conformal ghosts
and (at quadratic order) the longitudinal bosonic coordinates,
we will find that
determinants from these integrals will formally cancel each other
leaving only the partition function for the transverse coordinates.

	Consider the longitudinal
$(\xi^{0}, \xi^{1})$ part of
\quadcoord\ and integrate it by parts to put the lagrangian in the form
${\cal L} = \xi^{m}D_{mn}\xi^{n}$.  The result is
\eqn\ibpbos{
	\eqalign{
	S_{{\rm longitudinal}} =
		\frac{i}{2} \int \vol2 \biggl\{ & - \eta_{mn} \xi^{m}
	\left( 2\dem\dep - \frac{\dem\dep f}{f^{2}} + \frac{\dem f \dep f}
		{2f^{2}} \right) \xi^{n} \cr
		& \qquad \qquad \qquad \qquad - \epsilon_{mn} \xi^{m}
		\left( \frac{1}{f}\epsilon_{\alpha\beta}
		\d_{\alpha}f\d_{\beta}
		\right)\xi^{n} \biggr\}\ ,
}}
where $m$ and $n$ are restricted to 0 and 1 (which the {vielbeins}
map back to the $x-y$ plane).
Integrating over $\xi^{0}$ and $\xi^{1}$ leads to the determinant of the
above quadratic operator.
On the other hand,
Polyakov\ref\polyak{A.M. Polyakov \journal\pl&103B (81) 207}
showed that in writing an arbitrary metric as
\eqn\metricvar{
	h_{\alpha\beta}= \hat{h}_{\alpha\beta} \delta\phi
	+ \nabla_{\alpha}\varepsilon_{\beta}
	+ \nabla_{\beta}\varepsilon_{\alpha}\ ,
}
where $\nabla$ is the worldsheet covariant derivative,
the functional integration measure is
\eqn\metmeas{
	{\cal D}g_{\alpha\beta} = {\cal D}\phi
	{\cal D}\varepsilon_{\alpha}\det[L_{\alpha\beta}]\ ,
}
with
\eqn\diffop{
	L_{\alpha\beta} = \nabla_{\beta} \nabla_{\alpha} +
	\hat{h}_{\alpha\beta}
	\nabla^{\lambda}\nabla_{\lambda} - \nabla_{\alpha}\nabla_{\beta}
}
If we change variables from the coordinate
vectors $\varepsilon_{\alpha}$ to the
tangent space vectors $\varepsilon^{a}~=~e_{\alpha}^{a}\varepsilon^{\alpha}$,
where $e_{\alpha}^{a}$ is the {zweibein} for the worldsheet with metric
$\hat{h}_{\alpha\beta}$, then we get \metmeas\ with the
indices on the right hand side changed to tangent space indices and the
tangent space operator $L_{ab}$ identical to the operator in \ibpbos, so
that the Fadeev-Popov determinant formally cancels the
determinant arising from
integrating out the longitudinal modes.

	In performing the normal coordinate expansion, the fields became
contravariant vectors in spacetime, and so the kinetic
terms were endowed with a spin connection induced by the spacetime geometry.
This allowed us to use the determinants which arose from integrating
out the longitudinal fluctuations to cancel the Fadeev-Popov determinant
written with worldsheet covariant derivatives acting on worldsheet
vectors and spinors; the worldsheet metric is pulled back from the
spacetime metric, so it is reasonable to state that the worldsheet tangent
space structure, and thus the spin connection, is induced by the
spacetime tangent space structure.  Thus this cancellation is a
consequence of our choice of equilibrium classical solution and of
the particluar choice of fiducial metric shown in equation~\setfiducial.

It is important to keep straight in the above argument the distinction
between the intrinsic metric $h_{\alpha\beta}$ and the induced metric
$\gamma_{\alpha\beta}=\d_\alpha X^\mu\d_\beta X^\nu G_{\mu\nu}(X)$.
The determinants that arise at quadratic order in the functional
integral have the structure
\eqn\dets{{\rm det}_{\sst FP} [L_{ab}(h)]\
{\rm det}^{-1}_{\sst X^\pm} [L_{mn}(\gamma)]\
{\rm det}^{-1}_{\sst X^\perp} [\Delta_0(G)]
}
where $\Delta_0$ is the quadratic fluctuation operator of the
transverse modes.  The arguments of the differential operators
indicate their explicit dependences; of course each determinant
implicitly depends on $h$ through the regulator.
Since we are in the critical dimension, the
{\it derivative} of this product of determinants with respect
to $h$ vanishes.  On the other hand, if we are interested in
the {\it value} of the determinants at the point $h=\gamma$,
then the above treatment shows that the first two determinants
cancel and the partition function will be the determinant of the
transverse fluctuation operator regulated with the induced metric.

\subsec{Operator quantization, vacuum states, and Bogolubov transformations}

In constructing the Hilbert space of the transverse fluctuations we are
faced with the well-known problem of finding a reasonable physical vacuum
state (see for example \birrell~for a general discussion) given curved
worldsheet and spacetime manifolds.
In the absence of an exact quantization we are restricted to constructing
this vacuum perturbatively in the normal coordinate expansion.
Based on the analysis of the previous section,
this expansion is consistent with the solution of the Virasoro constraints
in conformal gauge.
Moreover we expect to be able to dress any
state in the Hilbert space of the transverse modes into a physical
state whose dependence on the longitudinal coordinates
is a small deviation
from the classical identification $\sigma^{-} = X^{x},
\sigma^{+} = X^{y}$ which may be corrected for systematically
(order by order).
In particular, we may use the classical
worldsheet~$\leftrightarrow$~spacetime correspondence to ask
what worldsheet fluctuations might be positive frequency with respect
to a spacetime observer.  Here we will assume that the aforementioned solution
to the Virasoro constraints may be arrived at quantum mechanically
in fixing to a physical, light-cone-like gauge along the lines of
section 3.1.
The problem then reduces, in lowest order, to
that of considering the transverse fluctuations in \quadcoord~as
a field theory in a 2d spacetime corresponding to the $r$~-~$t$ or
$x$~-~$y$ plane of a Schwarzschild black hole; scalar field theories
in this 2-D spacetime have been well
studied (see for example \unruh\birrell and references therein).

	We wish to assume that the ``in'' vacuum consists of no
excitations in the far past in Schwarzschild time.  On the
Kruskal manifold, this corresponds to asking that there are no field
excitations leaving $H^{-}$ and $\scrim$, in region I of
figure 1, and no excitations leaving $H^{+}$ and $\scrip$ in
region II of figure 1.
An observer at $\scrim_{I}$, living in an asymptotically flat region,
will be a Schwarzschild observer and thus will define positive frequency
modes with respect to the Schwarzschild time $t$.  An observer
leaving or falling into the horizon, or at least one for whom the
horizon is a finite time away,
will be a Kruskal
observer and thus will define positive frequency modes with respect
to a characteristic
Kruskal time; in particular, on $H^{\pm}$ and $\cal{I}^{\pm}$, the
timelike Killing vector becomes
lightlike so that the characteristic Kruskal time
will be $\sigma^{+}$ or $\sigma^{-}$ depending on the surface.
Thus, the ``in'' vacuum
is in region I annihilated by left-moving modes which are
positive-frequency with respect to $t$ and right-moving modes which are
positive frequency with respect to Kruskal time.  We are only asking
questions about observers in region I of our Kruskal diagram,  so we
have chosen the modes in region II for mathematical convenience --
phenomena in region I will not depend on how we treat region II.
This vacuum is the ``Unruh vacuum,'' used to mock up gravitational
collapse~\unruh\birrell\ref\fulling{S.A. Fulling
\journal{\it J. Phys. A: Math. Gen.}&10 (77) 917}.  We might use other
vacuums, for example by asking that there is no
radiation entering or leaving the horizon $H^{-}_{I}\cup H^{+}_{I}$ --
the ``Hartle-Hawking vacuum''~\birrell\ -- but the Unruh
vacuum is physically compelling and at any rate
the extension of our calculations to the Kruskal vacuum
is not difficult.

	The ``out'' vacuum we choose corresponds to an observer
in the far future ($\scrip_{I}$) who, living in an asymptotically flat
region, will define
positive frequency modes in terms of Schwarzschild time; thus the
``out'' vacuum is that annihilated by all modes which are positive frequency
with respect to $t$.  This is known as the ``Schwarzschild vacuum.''

For the right-moving modes used to define the Unruh vacuum
we use the solutions $(\sigma^{-})^{i\omega}$ which are, as we have stated
below \branch, are positive frequency for all (real) $\omega$.
Negative frequency modes are defined by taking the complex conjugate,
which is most easily computed by
writing $x^{i \omega}$ properly in terms of Schwarzschild modes
in region I and II,
\eqn\krtosch{
	\eqalign{
	x^{i a \omega} & =
	e^{\pi \omega a}\ e^{-i \omega u_{I}}\ \theta(-x) +
	e^{-i \omega u_{II}}\ \theta(x)\cr
	y^{1\omega a} & =
	e^{i\omega v_{I}} \theta(y) + e^{\pi\omega a} e^{i \omega v_{II}}
	\theta(-y) \cr
	a & = \frac{2r_{h}}{D-3}\ ,
}}
where $u_{I}$ and $u_{II}$ are the Schwarzschild null coordinates in
region I and II, respectively, of figure I, and $\theta$ is the
Heaviside function.  Note that due to the real exponential factor,
the complex conjugate of $x^{i\omega}$ is different from $x^{-i\omega}$.
More formally, complex conjugation will move the logarithmic branch cut
in $x^{i\omega}$ from the upper half plane as defined in \branch~to the
lower half plane, so that the complex conjugate is upper half plane analytic
and thus is composed of negative frequency modes.

	The transverse fluctuation operators may now be written in terms
of the two different mode decompositions.  The frequency-dependent
coefficients in the sums over modes arise from requiring that the
single-particle wavefunction in front of each operator is
normalized to a delta function under the Klein-Gordon norm \birrell, \unruh.
These decompositions are:
\eqn\krumodes{
	\eqalign{
	\xi^{k} = & \int^{\infty}_{0} \frac{d\omega}{\sqrt{4\pi|\omega|}}
	\left[ e^{-i\omega v_{I}} a_{I}^{k}(\omega) +
	e^{i\omega v_{I}} a_{I}^{k\dagger}(\omega)\right] \theta(y) \cr
	+ & \int_{-\infty}^{\infty} \frac{d\omega}{\sqrt{4\pi|\omega|
		\sinh(2\pi|\omega|)}} e^{-\half \pi\omega a}
		\left[ x^{i\omega a} b^{k}(\omega)
	+ (x^{i\omega a})^{\ast} b^{k\dagger}(\omega)\right] \cr
	+ & \int_{0}^{\infty} \frac{d\omega}{\sqrt{4\pi|\omega|}}
	\left[e^{i\omega v_{2}} a_{II}^{k}(\omega) + e^{-i\omega v_{II}}
		a_{II}^{k\dagger}(\omega)\right] \theta(-y)
}}
in the modes by which we define the Unruh vacuum, and
\eqn\schwamodes{
	\eqalign{
	\xi^{k} = & \int^{\infty}_{0} \frac{d\omega}{\sqrt{4\pi|\omega|}}
	\biggl[ \left( e^{-i\omega v_{I}}
		a_{I}^{k}(\omega) + e^{i\omega v_{I}}
		a_{I}^{k\dagger}(\omega)\right) \theta(y)\cr
	+ & \left( e^{-i\omega u_{I}} c_{I}^{k}(\omega) +
		e^{i\omega u_{I}} c_{I}^{k\dagger}(\omega)\right)
		\theta(-x) \cr
	+ & \left( e^{i\omega u_{II}} c_{II}^{k}(\omega) +
		e^{-i\omega u_{II}} c_{II}^{k\dagger}(\omega) \right)
		\theta(x) \cr
	+ & \left( e^{i\omega v_{II}} a_{II}^{k}(\omega) +
		e^{-i\omega v_{II}} a_{II}^{k\dagger}(\omega) \right)
		\theta(-y) \biggr]
}}
in the modes by which we define the Schwarzschild vacuum.
The operators $a_{I,II}, b, c_{I,II}$ each satisfy
canonical commutation relations
\eqn\comrel{
	[a_{I}^{k}(\omega), a_{I}^{k'\dagger}(\omega')] = \delta^{kk'}
	\delta(\omega - \omega')\ .
}
The modes describing left-moving excitations, created by $a_{I}^{k\dagger}$
and $a_{II}^{k\dagger}$ are the same for both decompositions so that
the only difference is in our definition of left-moving excitations.
The Bogolubov transformations between the $b$-operators defining the
Unruh vacuum and the $c_{I,II}$-operators defining the Schwarzschild
vacuum (we shall call these ``Kruskal'' and ``Schwarzschild'' operators,
respectively) are easily found by expanding $x^{i\omega a}$ as in \krtosch,
and are~\birrell:
\eqn\bogol{
	\eqalign{
	c_{I}^{k}(\omega) & = \frac{e^{\half \pi \omega	a}\ b_{I}^{k}(\omega)
		+ e^{-\half\pi\omega a} b^{k\dagger}(-\omega)}
	{\sqrt{2\sinh(\pi|\omega| a)}} \cr
	c_{II}^{k}(\omega) & =
	\frac{e^{\half\pi\omega a}\ b^{k}(-\omega) +
	e^{-\half\pi\omega a}\ b^{k\dagger}(\omega)}
	{\sqrt{2\sinh(\pi|\omega| a)}} \cr
	\omega & \geq 0\ .
}}

	We shall use these transformations in the next section
to compute the Hawking radiation of the black hole into
spatial fluctuations of the string.

\subsec{Physical states and Hawking radiation}

	One of the original motivations of this work was to figure out
whether the Virasoro constraints could be consistently imposed in
the covariant formalism
in the presence of Hawking radiation: one might think that
the Hawking radiation would populate all modes -- ghosts, longitudinal
modes, \etc\ -- thermally and so there is no way to guarantee that
an asymptotic observer would not see unphysical modes radiated by the black
hole.  However, in order to ask physical questions,
one must either work in a physical gauge or
work with physical operators in a covariant gauge, and define the
vacuum in a physical fashion.

	The first question that should be asked is whether
one may legitimately fix to the physical gauge we have chosen, or
in covariant language, whether one may consistently impose
the BRST condition
  $$ \qbrs \ket{{\rm physical}} = 0\ $$
and find an appropriate representation for the BRST cohomology classes.
We have not been able to rigorously
prove either, but given that classically one may fix a physical
gauge; that the theory is anomaly free to one loop (since the one-loop
\sig-model beta functional vanishes); and that the longitudinal
vacuum fluctuations explicitly cancel off the ghost vacuum fluctuations
in the partition function (again, to one loop), it seems that
we have sufficient
circumstantial evidence that the constraints may be solved explicitly
and consistently.

	Given a consistent physical gauge, we may simply work
with the transverse oscillators and solve for the longitudinal
oscillators order by order.  This is fine for computing quantities
such the size of transverse fluctuations $\langle \xi^{2} \rangle$ or
the expectation value of the transverse stress tensor, at least to lowest
order in the normal coordinate expansion, as we shall below.  Alternatively,
if BRST quantization is consistent, we expect that we can represent
the BRST cohomology classes by building states with
DDF-like operators\ref\ddf{E. Del Giudice, P. Di Vecchia and S. Fubini
\journal\anp&70 (72) 378}\ref\brower{R.C. Brower
\journal\pr&D6 (72) 1655}\ref\kato{M. Kato and K. Ogawa
\journal\np&B212 (83) 443.}\ref\thorn{C.B. Thorn
\journal\np&B286 (87) 61.}.
This also amounts to dressing the transverse oscillators with
longitudinal modes to make a physical state.
In building such states we are faced
with questions of what a physical (spacetime) observer sees as positive
frequency; the worldsheet coordinates have no meaning in covariant gauge,
so in particular the question of what is a positive
frequency mode on the worldsheet has no inherent spacetime meaning.
However, since the frequency of a DDF operator is a spacetime quantity,
we use it to define the vacuum directly with these manifestly physical
operators.  (This is exactly what Schoutens \et\
\ref\schout{K. Schoutens, H. Verlinde and E. Verlinde \journal\pr&D48
(93) 2670; Princeton preprint PUPT-1395, hep-th/9304128}
have done for their model of 2-D black hole evaporation.)
We see why our
original fear was naive; a proper formulation of the theory
avoids any question of unphysical modes before Hawking radiation ever
becomes an issue.

	This argument can be recast in terms of more physical
or heuristic descriptions of Hawking radiation.  If one considers
Hawking radiation as due to pair production near the horizon \birrell\
then the argument means essentially that
pair production involves only physical modes,
so that only physical states are radiated to $\scrip$.
Similarly, if one describes Hawking radiation via a density matrix
due to a summation over states at or behind the horizon, hidden
from our asymptotic observer at ${\cal I}^{+}$, then one will sum only
over states in the physical Hilbert space -- the full Hilbert space including
negative norm states being in some sense a fiction arising from the
gauge freedom remaining after passing to a covariant gauge.

\newsec{Physical manifestations of the Hawking process}

	Having discussed quantization and the difference between the ``in''
and ``out'' vacuums, one may begin to ask some physical questions about the
string.  The Hawking radiation in the 2d worldsheet field theory will
manifest itself in some thermal population of the transverse modes in the
``out'' basis, which means that an asymptotic (Schwarzschild) observer
will see the string fluctuating as if it was thermally excited with the
worldsheet temperature equal to the Hawking temperature
\eqn\hawktemp{
	k_{B} T_{H} = \frac{\hbar c (D-3)}{2\pi r_{h}}\ .
}
The three most apparent (to us) detectable manifestations of this radiation
are the thermal wandering of the string, the energy that is radiated
in string fluctuations out of the black hole towards ${\cal I}^{+}$, and
the production of physical spacetime modes
(microscopic strings) due to transitions
of the excited string.
We discuss the first two of these for the bosonic string
in the following subsections.

\subsec{Wandering of the string in the Schwarzschild background}

	As an estimation of the wandering of the string, we may calculate the
mean square deviation from the classical background solution,
\eqn\rmsdev{
	\langle \xi^{2}\rangle =
	\lim_{x\rightarrow x',y\rightarrow y'}\sum_{k=2}^{D-1}
	\bra{\Psi}:\xi^{k}(x,y)\xi^{k}(x',y'):\ket{\Psi}\ .
}
We need to fix both the state $\ket{\Psi}$ and the normal ordering
prescription.  $\ket{\Psi}$ we take to be the Unruh vacuum, but
we wish to normal order this expression with respect to the Schwarzschild
modes, on the assumption that normal ordering is a function of the
observer.  This means that we begin by expanding the operators $\xi^{k}$
according to~\schwamodes, pushing the Schwarzschild creation operators
to the left of the destruction operators, and then using~\bogol\ to
expand out $c^{k}$ in terms of $b^{k}$, so as to be able to
calculate the expectation value in the ``in'' state which is
annihilated by $a^{k}$ and $b^{k}$.  The normal ordered correlator is
\eqn\flucdiv{
	\langle :\xi(x,y) \xi(x',y'):\rangle =
	\frac{(D - 2) \ell_s^{2}}{4\pi} \int^{\infty}_{0}
	\frac{d\omega 2 \cos[\omega(u-u')]}
		{\omega\left(e^{2\pi\omega a} - 1\right)}\ ,
}
where we have put in a factor of
$\ell_s^{2} = \hbar^2 c^2 2\pi\alpha' = \hbar c/T$, where $T$ is
the string tension
\ref\goddard{P. Goddard, J. Goldstone, C. Rebbi and C.B. Thorn
\journal\np&B56 (73) 109},
to give $\langle \xi^{2}\rangle$ the correct dimensions.  We note immediately
that the normal ordering has removed the ultraviolet divergences but
a nasty infrared divergence remains.  Eq. \flucdiv\ has the form
$\int d\omega D(\omega) \langle n(\omega, T_{H})\rangle$ (with $D$
being the density of states) of a 2d Bose gas of $D-2$ massless particles,
where $k_{B} T_{H} = 1/\beta = 1/(2\pi a)$.
The IR divergences arising from the density of states and
from the Planck distribution function are characteristic of this
general statistical mechanical system in the infinite volume limit, and are
entirely physical.
Also note that the transverse stress tensor $\d \xi \d \xi$
is infrared finite at this order.
To make sense of \flucdiv, we need a physically
motivated finite-size or finite-time cutoff -- an example of the latter
is the finite lifetime of the black hole.

Thus if one regulates~\flucdiv\ with an infrared momentum cutoff at momentum
$\epsilon$, one finds that $\langle \xi^{2} \rangle$ diverges linearly
with the cutoff,
\eqn\brcut{
	\langle \xi^{2} \rangle \sim
		\frac{(D-2)\ell_s^{2}}{2\pi\beta\epsilon}\ .
}
The deviation then goes as the square root of the finite-size or finite-time
cutoff $1/\epsilon$, so that the average string configuration is that of a
random walk.  This is a result
we might have expected from studies of average string configurations
at finite temperature;
in particular, Mitchell and Turok \ref\mitch{D. Mitchell and
N. Turok \journal\prl&58 (87) 1577; \andjournal\np&B294
(87) 1138} found that for closed bosonic strings at finite
temperature, the
mean square radius of the string scaled as the length $L$ of the string,
$$ \langle \Delta r^{2} \rangle \sim L\ .$$
Nevertheless the angular variation $\delta\theta\sim \delta\xi/r$
is well-behaved, so although the string fluctuates far from its
center of mass, it does not cover a large solid angle of the black hole.

The presence of this
infrared catastrophe indicates that although the equations of motion
derived from \quadact\ are stable, we expect that quantum fluctuations
will require us to include some nonlinear effects
in the worldsheet effective action.
One might worry that, since the thermal propagator for the transverse
string fluctuations is quadratically infrared divergent, that higher
order corrections to \eg\ the stress tensor will be much larger than
the effects we have calculated (see below).
We believe that what is breaking down
is not the order of magnitude of the radiated Hawking flux, but rather
the Taylor expansion of certain quantities in the action.
For example, part of the small fluctuation expansion involves the
Taylor expansion of terms
like $G(X_{\rm cl}+\xi)\partial\xi\partial\xi$, which has
a radius of convergence of order $r_{\rm cl}$.
So instead of treating  perturbatively the quantum fluctuations of all the
modes, we should introduce an infrared cutoff and treat the
low frequency oscillations adiabatically and classically (since they
have large occupation numbers) while quantizing all the higher frequency
modes.  This scheme should work quite well if we stay far
from the horizon.

\subsec{Energy radiated onto the string}

	We would like some idea of the energy radiated onto the string in
the Hawking process.  The world-sheet stress tensor measures the
world-sheet energy radiated out along the string, so a definite
relation between world-sheet and spacetime time coordinates will
relate world-sheet and spacetime energy densities and fluxes.
Such a relation is provided by the classical solution \trivsoln --
hence we expect that the spacetime and world-sheet energy fluxes
should coincide.

	We shall work entirely in the asymptotically flat
regime of spacetime, in Schwartzchild null coordinates.
In this region the Lagrangian~\sigact\ is manifestly
(approximately) Lorentz invariant and so there is an asymptotically
well defined Noether current
\eqn\momcur{
	P_{\alpha \mu} = T g_{\mu\nu}\d_{\alpha}X^{\nu}\ ,
}
where $T$ is the string tension.
The total momentum radiated to future infinity will then be
\eqn\gonemom{
	P_{\mu}^{{\rm rad}} = \int_{\cal{I}^{+}} d\sm P_{-\mu}\ .
}
Expanding in normal coordinates, we find that
\eqn\pexp{
	P_{\alpha\mu} = P_{({\rm cl})\alpha\mu} + \delta
		P_{\alpha\mu} =
	T (\d_{\alpha} X_{0}^{\mu} + D_{\alpha}
		\xi^{\mu} + {\ldots})\ .
}
Now $-P^{\mu}P_{\mu}$ should be the total mass-squared radiated away to
$\scrip$, so that
\eqn\massrad{
	\eqalign{
	-2 f P^{u} P^{v} + r^{2} \left(P^{\perp}\right)^{2}  & =
		- 2 f P^{u}_{c} P^{v}_{c} -
		2 f P^{u}_{c}\delta P^{v} -
		2 f P^{v}_{c}\delta P^{u} - \cr
		&\qquad\quad - f\delta P^{u} \delta P^{v}
		+ (P^{\perp}_{c})^{2} +
	2(P^{\perp}_{c}\cdot\delta P^{\perp}) +
	(\delta P^{\perp})^{2} \cr
	& = - (M_{c} + \delta M)^{2}\ .
}}
where the subscript $c$ denotes ``classical.''  We use the classical
mass shell condition and our explicit values for $X^{\mu}_{c}$
to write
\eqn\moremassrad{
	- 2 f P^{u}_{c} \delta P^{v} + (\delta P^{\perp})^{2}
		= - 2 M_{c}\delta M - \delta M^{2}\ .
}
The integrals in \gonemom\ used to define the total momentum $P^{\mu}$
will be infinite and will furthermore select out the
zero modes of $\d_{-} X^{\mu}$.  We thus find an expression for
mass per unit interval along $\scrip$:
\eqn\massdens{
	-2 f \d_{-}X_{c}^{u}(\d_{-}\xi^{v})_{0} +
		(\dem \xi^{\perp}_{0})^{2} + {\ldots}
	= - \frac{2M\delta{M}}{{(\rm length)^2}}
		- \left(\frac{\delta M}{{\rm length}}\right)^{2}\ ,
}
where the subscript $0$ indicates the space-independent part.  We will
drop the $\delta M^{2}$ term as being higher order in our expansion.
Substituting in the lowest order light-cone condition in
Schwarzchild null coordinates, we find that
$$ 2 f \d_{-} X^{u}_{c} \d_{-} \xi^{v} = T^{\perp}_{--} $$
(where here we have set $\sm = u$, $\sp = v$), and thus
\eqn\massapprox{
	T^{\perp}_{-- ,0} - (\dem \xi^{\perp}_{0})^{2} = \frac{1}{T^{2}}
	\frac{2M\delta M}{{(\rm length)^2}} .
}
The second term on the left hand side should vanish since we are looking
at oscillating transverse solutions with zero total transverse momentum.
Now ($M/{{\rm length}}$) is just the string tension, so that, converting
energy density to energy flux,
\eqn\flux{
	\frac{\delta M}{{\rm time}} = \frac{c}{4\pi\hbar c \alpha'}
		T_{--,0}^{\perp}
}
In order to compute the Hawking
radiation, we wish to take the vacuum expectation value of
$:T^{{\perp}}:$
in the Unruh vacuum, with normal ordering performed as before with respect
to the Schwarzchild vacuum.  With this vacuum
and normal ordering prescription, only $T^{\perp}_{--}$ will
be non-zero, and we find that to lowest order,
\eqn\transstennum{
	\langle :T^{\perp}_{--}: \rangle =
	(D-2) \frac{\pi \ell_s^{2}}{12 \beta^{2}}\ .
}
The mass per unit time radiated out onto the string is thus found to be:
\eqn\flux{
	\hbox{energy flux} = \frac{\delta M}{\hbox{time}}
		\sim \frac{\hbar c^2}{r_{h}^{2}}\ ,
}
where we have put all of the physical constants back in.

We may ask when the energy radiated onto the string
modes is larger than the radiation into spacetime modes.
The energy flux into, for example, a scalar field such as the
string dilaton is essentially the same as calculated above
since the radiation is mostly into S-wave states:
\eqn\targetflux{
	\PP \sim \frac{\hbar c^{2}}{r_{h}^{2}}\ .
}
Thus the energy flux of radiation onto string modes is proportional to
the energy flux of spacetime scalars.

In fact both the world-sheet and spacetime field energy fluxes differ
from the estimates \flux\ and \targetflux\ in several ways.
Both the spacetime field modes and the angular string fluctuations
will suffer backscattering from the gravitational field near the horizon.
This backscattering will be frequency-dependent, and for energies
typical of the Hawking radiation leads to a non-negligible fraction of
the radiation to be reabsorbed by the black hole.
Curiously, although the shape of this barrier is in general different
for spacetime S-waves and string angular fluctuations, the two
barriers are identical in $D=4$.
The spacetime field modes of nonzero angular momentum
also contribute some flux (decreasing exponentially in the angular
momentum) which does not significantly alter the total flux.
Finally, it should be emphasized that the string is generally
propagating in a spacetime $\MM_{\rm schw}^{D}\times\KK^{d_{cr}-D}$,
the product of the $D$-dimensional Schwarzschild geometry we have
been discussing and an unspecified internal manifold $\KK$
that together with $\MM_{\rm schw}$
makes up the total effective central charge $d_{cr}$
of the sigma model in which the string propagates.
The string coordinates of this internal space will also Hawking radiate,
and in this case there is no barrier from the Schwarzschild curvature.
The corrections to the flux
in internal coordinates come from the difference between
the density of states of the sigma
model on $\KK$ and that of free fields.
If the sigma model on $\KK$ is weakly coupled,
in the normal coordinate
expansion the corrections to the stress tensor
will be down by powers of $\ell_s/r_\KK$
where $r_\KK$ is the typical radius of curvature of $\KK$.
Thus the total flux in spacetime modes will be roughly proportional
to the number of massless spacetime fields (remembering to
count polarization states and correct for the angular momentum
barrier in case of intrinsic spin), while the flux in string modes will
be roughly proportional to $d_{cr}-2$.

Finally, one should remember that when the black hole entraps the
macroscopic string, two semi-infinite pieces protrude from the horizon.
This doubles the flux into string modes -- or does it?
For chirally asymmetric strings,
one of the two pieces propagates `left-moving' string modes
in the radially outward direction; the other piece
propagates `right-moving' string modes radially outward.  Thus
the energy-momentum flux in general differs for the two halves.
This effect is most pronounced for the heterotic string in the
critical dimension, where the left-moving (bosonic string)
flux is that of 24 bosons, while that of the right-moving
(fermionic string) flux is effectively that of 12 bosons
(\eg\ by bosonizing the eight transverse fermions).  Curiously,
the back-reaction of the Hawking radiation will not only
evaporate the black hole, but due to the asymmetric radiation
pressure will accelerate the black hole down the string!

\newsec{Discussion}

One issue we have not really settled is how to restrict the string
functional integral to the region of field space
outside the horizon.  In principle the string fluctuates away from
the classical solution in such a way that part of the string dips
into the black hole.
If we take seriously the ultraviolet fluctuations of the world-sheet,
these `dips' are ubiquitous (and indeed extend to the singularity),
although incoherent.
A bit of averaging over short distances replaces such fluctuations
by a renormalized wandering of order the renormalization scale.
We leave a detailed investigation of the
structure of physical states near the horizon to future work.

Another issue is the relation between the Bekenstein-Hawking
entropy as measured by the world sheet, in comparison to that measured
by spacetime physics.  The former will be either
$\log M$ or constant  as in the two-dimensional
model system of~\cghs, whereas the latter will be proportional to the
$(D-2)$-dimensional area of the event horizon; what states of the
black hole does the string have access to?  The naive calculation would
suggest only the `states near the horizon' where the string
is attached to the black hole.
It has been claimed recently that the divergent quantum corrections
to the entropy are cut off in string theory~\ref\lenny{L. Susskind,
{\it Some speculations about black hole entropy in string theory},
Rutgers preprint RU-93-44, hep-th/9309145.} because
of string theory's soft spacetime high energy behavior.
However string theory still has divergences on the world sheet,
and those presumably still contribute a logarithmic divergence to
the macroscopic string's entropy, unless there is a relation
between the world sheet cutoff and the spacetime cutoff
provided by the soft high-energy behavior of strings.  But how does such
a relation come about?  Again it is not clear.

A logical extension of the present work (currently being contemplated)
involves the construction of the physical vacuum.  According to
our discussion of section (4.3), this vacuum is not an incoherent
superposition of all modes, but rather of a
thermal distribution of transverse modes which are {\it coherently}
dressed by ghosts and longitudinal modes.  We intend to build such
a state out of DDF-style operators for the dressed transverse oscillation
modes, with the ultimate goal of calculating the amplitudes for emission
and scattering of microscopic strings.

One may also contemplate a number of other model calculations
which go beyond the scope of the present work.  One would like to
study not just the evaporation problem but also the formation of
black holes in the macroscopic string model, as in the two-dimensional
system of Callan \et\cghs.  This might be achieved by an
appropriate calculation of the gravitational field back-reaction
of a macroscopic string.  Naively the left- and right-moving modes of
a fundamental string in flat space are free fields.  However, we
expect two counterpropagating pulses (`shock waves') to interact
gravitationally, and if sufficiently energetic, to form a black
hole upon collision.  Generally one must also take into account the
other long-range fields carried by the string (\eg\ the axion and
dilaton charge per unit length carried by macroscopic fundamental
strings \ref\harvey{A. Dabholkar, G. Gibbons, J. Harvey and
F. Ruiz Ruiz\journal \np&B340 (90) 33.}).
The appropriate starting point for such a calculation
might be the family of exact string solutions of Garfinkle
\ref\garfinkle{D. Garfinkle\journal
\pr&D45 (92) 4286.}\ref\harvey2{J. Gauntlett,
J. Harvey, and D. Waldram, to appear.}.

The proper relation between the string and spacetime fields
were only partially addressed here.  It would be interesting
to exhibit a self-consistent world-sheet calculation that
would incorporate the
back-reaction of the macroscopic string dynamics on the spacetime
fields, for instance to show in a world-sheet beta function calculation
how the energy radiated along the string contributes to a decrease
in the black hole mass.
We imagine such effects should appear in a term of the Polyakov
type $(\partial \log[f(X)])^2$ in the world-sheet
effective action, which then couples to the spacetime effective
action in the manner described in \harvey.
This mechanism is somewhat confusing in that the
macroscopic string is a single quantum from the point of view of the
string field theory, and we do not ordinarily incorporate the
classical field of a single quantum in the classical background
(\eg\ we do not treat single electrons as small black holes
or background Coulomb charges).
Presumably one needs to resum the emission of soft gravitons from
the world sheet into an effective classical field, and such nearly
on-shell gravitons will give logarithmic almost-divergences in the
world-sheet theory which resum to a modified classical field potential.
{}From the worlsheet point of view these microscopic strings
that dress the spacetime geometry
are a resummation of `baby universe' processes.
It would be nice to understand the details.

\newsec{Acknowledgements}

We wish to thank J. Harvey, V. Iyer, D. Kutasov, and L. Susskind
for a number of discussions, explanations, and useful comments.

\listrefs
\bye